\documentclass[journal]{IEEEtran}

\usepackage[dvips]{color}
\usepackage{graphicx}
\usepackage{amsmath}

\begin{document}

\title{On Secrecy Performance of Antenna Selection Aided MIMO Systems Against Eavesdropping}

\markboth{IEEE Transactions on Vehicular Technology (accepted to appear)}%
{Jia Zhu \MakeLowercase{\textit{et al.}}: On Secrecy Performance of Antenna Selection Aided MIMO Systems Against Eavesdropping}

\author{Jia~Zhu, Yulong~Zou,~\IEEEmembership{Senior Member,~IEEE}, Gongpu~Wang, Yu-Dong Yao,~\IEEEmembership{Fellow,~IEEE},
and George K. Karagiannidis,~\IEEEmembership{Fellow,~IEEE}

\thanks{Copyright (c) 2013 IEEE. Personal use of this material is permitted. However, permission to use this material for any other purposes must be obtained from the IEEE by sending a request to pubs-permissions@ieee.org.}

\thanks{Manuscript received September 21, 2014; revised December 7, 2014 and January 18, 2015; accepted January 25, 2015. The editor coordinating the review of this paper and approving it for publication was Dr. Maged Elkashlan.}

\thanks{This work was partially supported by the Priority Academic Program Development of Jiangsu Higher Education Institutions, the National Natural Science Foundation of China (Grant Nos. 61302104 and 61401223), the Scientific Research Foundation of Nanjing University of Posts and Telecommunications (Grant Nos. NY213014 and NY214001), the 1311 Talent Program of Nanjing University of Posts and Telecommunications, and the Natural Science Foundation of Jiangsu Province (Grant No. BK20140887).}

\thanks{J. Zhu and Y. Zou (corresponding author) are with the School of Telecommunications and Information Engineering, Nanjing University of Posts and Telecommunications, Nanjing, Jiangsu 210003, China. (Email: \{jiazhu, yulong.zou\}@njupt.edu.cn)}

\thanks{G. Wang is with the School of Computer and Information Technology, Beijing Jiaotong University, Beijing, China. (Email: gpwang@bjtu.edu.cn)}

\thanks{Y.-D. Yao is with the ECE Department, Stevens Institute of Technology, New Jersey, USA. (Email: yyao@stevens.edu)}

\thanks{G. K. Karagiannidis is with the ECE Department, Aristotle University of Thessaloniki, Greece. (Email: geokarag@auth.gr)}

}

\maketitle

\begin{abstract}
In this paper, we consider a multiple-input multiple-output (MIMO) system consisting of one source, one destination and one eavesdropper, where each node is equipped with an arbitrary number of antennas. To improve the security of source-destination transmissions, we investigate the antenna selection at the source and propose the optimal antenna selection (OAS) and suboptimal antenna selection (SAS) schemes, depending on whether the source node has the global channel state information (CSI) of both the main link (from source to destination) and wiretap link (from source to eavesdropper). Also, the traditional space-time transmission (STT) is studied as a benchmark. We evaluate the secrecy performance of STT, SAS, and OAS schemes in terms of the probability of zero secrecy capacity. Furthermore, we examine the generalized secrecy diversity of STT, SAS, and OAS schemes through an asymptotic analysis of the probability of zero secrecy capacity, as the ratio between the average gains of the main and wiretap channels tends to infinity. This is different from the conventional secrecy diversity which assumes an infinite signal-to-noise ratio (SNR) received at the destination under the condition that the eavesdropper has a finite received SNR. It is shown that the generalized secrecy diversity orders of STT, SAS, and OAS schemes are the product of the number of antennas at source and destination. Additionally, numerical results show that the proposed OAS scheme strictly outperforms both the STT and SAS schemes in terms of the probability of zero secrecy capacity.

\end{abstract}


\IEEEpeerreviewmaketitle

\section{Introduction}

\IEEEPARstart {D}{ue} to the broadcast nature of wireless medium, the radio signal of a source node can be overheard by any unauthorized user within its transmit coverage, which makes the wireless communications vulnerable to eavesdropping attacks. In order to achieve secure wireless transmissions, cryptographic techniques have been widely used to ensure that the confidential information can be decoded by the legitimate receiver only while preventing an eavesdropper from the interception. In addition to the conventional cryptographic techniques, physical-layer security is now emerging as a new secure communication paradigm by exploiting the physical characteristics of wireless channels to prevent the eavesdropper from intercepting the information exchange between legitimate users.

Physical-layer security work was pioneered by Shannon [1] and later extended by Wyner [2], where an information-theoretic framework has been established by developing achievable secrecy rates for a classical three-node scenario consisting of source, destination and eavesdropper. More specifically, Wyner showed that when the main channel spanning from source to destination has a better conditional than the wiretap channel from source to eavesdropper, there exists a positive rate at which the source and destination can communicate reliably and securely. In [3], Wyner's results were further extended to the Gaussian wiretap channel, where the secrecy capacity is shown as the difference between the capacity of the main channel and that of the wiretap channel.

\subsection{Related Literature}
It is known that the wireless capacity is severely degraded due to the channel fading effect. To this end, multiple-input multiple-output (MIMO) was proposed as an effective means to combat wireless fading and improve channel capacity [4], [5], which also has great potential to increase the secrecy capacity of wireless transmissions and enhance the wireless physical-layer security. In [6] and [7], the secrecy capacity of multiple-input single-output (MISO) wiretap channel was examined and characterized in terms of generalized eigenvalues. In [8], the authors studied the maximal achievable rates of MIMO wiretap channel and proposed a generalized-singular-value-decomposition (GSVD) scheme to achieve the secrecy capacity. In [9], the MIMO broadcast wiretap channel was investigated from an information-theoretic perspective in terms of the secrecy capacity. Furthermore, in [10]-[12], the secrecy capacity of wiretap channel in wireless fading environments was examined and evaluated by using optimal power and rate allocation strategies. In addition, user cooperation [13]-[15] as virtual MIMO by allowing users to share each other's antennas was shown in [16] and [17] to improve the wireless secrecy capacity. For example, in [19], we studied cooperative relay selection for securing wireless communications and demonstrated that the wireless physical-layer security significantly improves with an increasing number of relays. In [20] and [21], multiuser scheduling was shown as a promising approach to protect cognitive radio networks against eavesdropping.

Recently, extensive efforts have been devoted to the research of transmit antenna selection for the wireless physical-layer security. In [22], the authors explored the transmit antenna selection for a MISO communication system consisting of a multiple-antenna transmitter and a single-antenna receiver in the presence of a multiple-antenna eavesdropper. It was shown in [22] that the transmit antenna selection considerably enhances the wireless security in terms of secrecy outage probability. Later on, the antenna selection work of [22] was further extended in [23] and [24] to a MIMO communication system comprised of a source, a destination and an eavesdropper, each equipped with multiple antennas. Closed-form expressions of the secrecy outage probability were derived in [23] and [24] for the transmit antenna selection assisted MIMO communications in fading environments. In [25], the impact of outdated channel state information (CSI) on the transmit antenna selection was examined for a MISO system, showing that the secrecy outage performance expectedly degrades when the CSI obtained at the transmitter is outdated due to the CSI feedback delay.

It is worth mentioning that all the aforementioned studies [22]-[25] assume only the CSI of the main channel available at the transmitter without knowing the eavesdropper's CSI knowledge. The transmit antenna selection with the global CSI of both the main channel and wiretap channel was analyzed in [26], which, however, is limited to the performance evaluation of a simple MISO system only in terms of the average secrecy rate. The performance of optimal antenna selection with the global CSI knowledge remains unknown for MIMO communication systems, which will provide a theoretical upper bound as a guide for developing new approaches to defend against eavesdropping attacks. Additionally, in existing literature (e.g. [23], [24], [27] and [29]), the secrecy diversity was established and analyzed by assuming that the average received signal-to-noise ratio (SNR) at the destination tends to infinity, while the eavesdropper's received SNR is finite. As an alternative, we presented a generalized secrecy diversity definition in [19]-[21] for characterizing an asymptotic behavior on the secrecy performance, as the ratio between the average gains of the main channel and the eavesdropper's wiretap channel approaches infinity, called main-to-eavesdropper (MER) ratio. To the best of our knowledge, it is still an open issue to examine our generalized secrecy diversity for the transmit antenna selection assisted MIMO communications in the face of an eavesdropper.

\subsection{Motivation and Contribution}
{As aforementioned, no previous work addresses the optimal antenna selection with the global CSI knowledge of both the main channel and wiretap channel for MIMO communication systems, where the source, destination and eavesdropper are equipped with an arbitrary number of antennas. It has been shown in [3] and [9] that the secrecy capacity of MIMO communications is given by the difference between the capacity of the main channel and that of the wiretap channel. To be specific, when the main channel has a better quality than the wiretap channel, a positive secrecy capacity is achieved and the source can transmit at a non-zero rate to the destination reliably and securely. However, if the main channel is a degraded version of the wiretap channel, then the secrecy capacity becomes zero and an event of zero secrecy capacity occurs. In this case, the source and destination cannot transmit reliably and securely, which is referred to as an intercept event in [19]. This paper is focused on deriving analytical expressions of the probability of zero secrecy capacity for transmit antenna selection aided MIMO communications in Rayleigh fading environments. Also, we examine the generalized secrecy diversity [19]-[21] for MIMO communications through an asymptotic analysis of the probability of zero secrecy capacity in high MER region, which provides a useful insight into the effect of the number of antennas on wireless security against eavesdropping.} The main contributions of the paper are summarized as follow.

\begin{itemize}
\item
{We explore the transmit antenna selection for physical-layer security of a MIMO system consisting of one source and one destination in the presence of an eavesdropper, where each node is equipped with an arbitrary number of antennas. The optimal antenna selection (OAS) and suboptimal antenna selection (SAS) schemes are proposed to improve the security of source-destination transmissions against the eavesdropper. Specifically, in the OAS scheme, the source is assumed to have the global CSI knowledge of the main channel and wiretap channel, whereas only the main channel's CSI is known in the SAS scheme without requiring the eavesdropper's CSI. As discussed above, the existing antenna selection work (see e.g. [22]-[25]) considers that only the main channel's CSI is available at the source node and, moreover, the optimal antenna selection with the global CSI knowledge studied in [26] is constrained to a MISO scenario without considering a more general MIMO system. For comparison purposes, we also consider the conventional space-time transmission (STT) as a benchmark.}
\end{itemize}

\begin{itemize}
\item
We consider the probability of zero secrecy capacity as a metric to evaluate the security performance of STT, SAS and OAS schemes. {{To be specific, an event of the zero secrecy capacity occurs when the capacity of the main channel spanning from the source to the destination falls below the capacity of the wiretap channel from the source to the eavesdropper. In this case, the eavesdropper would be able to succeed in intercepting the source-destination transmission. The probability of zero secrecy capacity is thus to evaluate the probability that the physical-layer security cannot be achieved.}} We derive analytical expressions of the probability of zero secrecy capacity for the STT, SAS and OAS schemes. It is shown that the probabilities of zero secrecy capacity for the STT, SAS and OAS schemes are independent of the SNR, implying that increasing the transmit power cannot improve the wireless security in terms of the probability of zero secrecy capacity.
\end{itemize}

\begin{itemize}
\item
{We examine the use of our generalized secrecy diversity [19]-[21] to characterize an asymptotic behavior on the probability of zero secrecy capacity, as the MER tends to infinity. It needs to be pointed out that as aforementioned, the probability of zero secrecy capacity is independent of the SNR, which makes the conventional SNR based secrecy diversity (e.g. [23], [24], [27] and [29]) becomes not applicable here. By contrast, our generalized secrecy diversity attempts to show an asymptotic probability of zero secrecy capacity in high MER region, which is motivated by the fact that the probability of zero secrecy capacity is mainly determined by the average gains of main and wiretap channels. Using our generalized secrecy diversity definition, we obtain that the STT, SAS and OAS schemes achieve the full diversity order of $MN_d$, where $M$ and $N_d$ represent the number of antennas at source and destination, respectively. This coincidentally matches the conventional secrecy diversity result of [23]. It is also implied that the secrecy diversity performance of STT, SAS and OAS schemes is independent of the number of eavesdropper's antennas.}
\end{itemize}

\subsection{Organization and Structure}
The remainder of this paper is organized as follows. Section II describes the system model and presents the STT, SAS, and OAS schemes. {{Next, we conduct the performance analysis of the STT, SAS and OAS schemes in terms of the probability of zero secrecy capacity and the secrecy diversity in Section III, followed by Section IV}}, where numerical results are evaluated to show the security advantage of proposed antenna selection over conventional space-time coding. Finally, we make some concluding remarks in Section V.

\section{System Model and Problem Formulation}
\begin{figure}
  \centering
  {\includegraphics[scale=0.6]{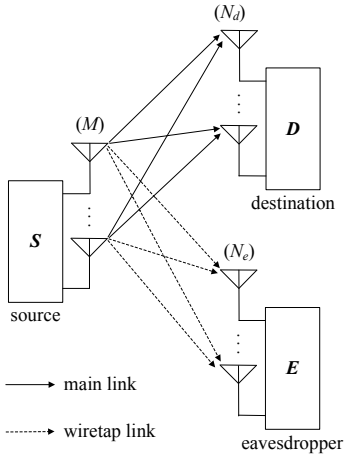}\\
  \caption{A multiple-input multiple-output (MIMO) system consisting of one source (\emph{s}) and one destination (\emph{d}) in the presence of an eavesdropper (\emph{e}).}\label{Fig1}}
\end{figure}

As shown in Fig. 1, we consider a wireless MIMO system consisting of a source (\emph{s}) and a destination (\emph{d}) in the presence of an eavesdropper (\emph{e}),
in which the solid and dash lines represent the main channel (from source to destination) and the wiretap link (from source to eavesdropper), respectively. In Fig. 1, each network node is equipped with an arbitrary number of antennas, where $M$, $N_d$ and $N_e$ represent the number of antennas at the source, destination and eavesdropper, respectively. Moreover, the sets of antennas at the source, destination and eavesdropper are denoted by $\mathcal{S}$, $\mathcal{D}$ and $\mathcal{E}$, respectively.
Additionally, both the main and wiretap channels are modeled as the Rayleigh fading and the thermal noise received at any node is modeled as a zero-mean complex Gaussian random variable (RV) with a variance of $\sigma^2_n$, i.e., ${\mathcal{CN}}(0,\sigma^2_n)$. {Although only the Rayleigh fading model is considered in this paper, similar performance analysis and results could be obtained for other fading models e.g. Nakagami fading.}

In MIMO communications systems, space-time coding [30] has been widely recognized as an effective means to improve the performance of wireless transmissions. In what follows, we first consider the space-time transmission (STT) as a benchmark and then examine the transmit antenna selection to improve the physical-layer security for MIMO systems. We present two antenna selection schemes, namely the optimal antenna selection (OAS) and suboptimal antenna selection (SAS), which operate depending on whether or not the global CSI of both the main and wiretap channels is available at the source. To be specific, in the OAS scheme, the global CSI is known at the source, whereas only the main channel's CSI is assumed in the SAS scheme without knowing eavesdropper's CSI.

\subsection{STT Scheme}
In the STT scheme, the source signal is first encoded by using space-time coding [30]. Then, all $M$ transmit antennas participate in sending the encoded signal to the destination, where the total amount of transmit power across $M$ antennas is constrained to a fixed value i.e. $P$. For simplicity, we here consider an equal-power allocation method, leading to the transmit power of $P/M$ for each transmit antenna. Assuming that the destination has the perfect CSI of the main channel for space-time decoding [30], the receive signal-to-noise ratio (SNR) at the destination relying on the STT scheme can be given by
\begin{equation}\label{equa1}
\gamma^{{\textrm{STT}}}_d=\sum\limits_{i = 1}^M {\sum\limits_{j = 1}^{{N_d}} {\frac{{P|{h_{id_j}}{|^2}}}{M{\sigma^2_n}}} },
\end{equation}
where $h_{id_j}$ is a fading coefficient of the main channel from transmit antenna $i$ of the source to receive antenna $j$ of the destination. Thus, according to (1), the maximal achievable rate of the source-to-destination transmission relying on the STT scheme is expressed as
\begin{equation}\label{equa2}
R^{\textrm{STT}}_{sd}=\log_2(1+\gamma^{{\textrm{STT}}}_d).
\end{equation}
Similarly, the maximal achievable rate at the eavesdropper with STT scheme can be given by
\begin{equation}\label{equa3}
R^{\textrm{STT}}_{se}=\log_2(1+\gamma^{{\textrm{STT}}}_e),
\end{equation}
where $\gamma^{{\textrm{STT}}}_e$ represents the receive SNR at the eavesdropper as given by
\begin{equation}\label{equa4}
\gamma^{{\textrm{STT}}}_e=\sum\limits_{i = 1}^M {\sum\limits_{j = 1}^{{N_e}} {\frac{{P|{h_{ie_j}}{|^2}}}{M{\sigma^2_n}}} },
\end{equation}
where $h_{ie_j}$ is a fading coefficient of the wiretap channel from transmit antenna $i$ to receive antenna $j$ of the eavesdropper.

\subsection{OAS Scheme}
In the OAS scheme, only the ``best" transmit antenna will be selected and used for transmitting the source signal to the destination. We here assume that the source node has the global CSI of both the main and wiretap channels, where {the CSI of wiretap channel may be estimated and obtained by monitoring the eavesdropper's transmissions [12], [17]. Notice that the eavesdropper could be a legitimate user who is interested in tapping other users' signals, which may be active in the network.} It is pointed out that the following subsection will address the scenario, where the wiretap channel's CSI is unavailable at the source. Without loss of generality, let us consider that the transmit antenna $i$ is selected as the ``best" antenna to transmit the source signal with power $P$. Considering the use of maximal ratio combining (MRC) at the destination, we can obtain the maximal achievable rate of the transmission from transmit antenna $i$ to destination as
\begin{equation}\label{equa5}
R_{id}=\log_2(1+\sum\limits_{j = 1}^{{N_d}} {\frac{{P|{h_{id_j}}{|^2}}}{{\sigma _n^2}}}).
\end{equation}
Similarly, the maximal achievable rate obtained at the eavesdropper is given by
\begin{equation}\label{equa6}
R_{ie}=\log_2(1+\sum\limits_{j = 1}^{{N_e}} {\frac{{P|{h_{ie_j}}{|^2}}}{{\sigma _n^2}}}).
\end{equation}
Obviously, the transmit antenna that maximizes the secrecy capacity can be viewed as the ``best" antenna. As discussed in [3] and [9], the secrecy capacity is given by the difference between the capacity of the main channel and that of the wiretap channel. Thus, the optimal antenna selection criterion can be given by
\begin{equation}\label{equa7}
\begin{split}
b = \arg \mathop {\max }\limits_{i \in {\mathcal{S}}} ({R_{id}} - {R_{ie}})=\arg \mathop {\max }\limits_{i \in {\mathcal{S}}} \dfrac{{1 + \sum\limits_{j = 1}^{{N_d}} {\dfrac{{P|{h_{id_j}}{|^2}}}{{\sigma _n^2}}} }}{{1 + \sum\limits_{j = 1}^{{N_e}} {\dfrac{{P|{h_{ie_j}}{|^2}}}{{\sigma _n^2}}} }},
\end{split}
\end{equation}
where $b$ denotes the ``best" antenna and $\mathcal{S}$ represents a set of $M$ transmit antennas at the source node. Note that the transmit power $P$ in (7) is known at the source. Moreover, the thermal noise variance $\sigma _n^2$ is shown as $\sigma _n^2 =\kappa TB$ [33], where $\kappa $ is the Boltzmann constant (i.e., $\kappa  = 1.38 \times {10^{ - 23}}{\textrm{\emph{Joule/Kelvin}}}$), $T$ is the room temperature in $\textrm{\emph{Kelvin}}$, and $B$ is the system bandwidth in $\textrm{\emph{Hz}}$. Since both the room temperature $T$ and system bandwidth $B$ are predetermined, the noise variance $\sigma_n^2$ can be easily obtained at the source. Therefore, once the global CSI of the main and wiretap channels $|{h_{id_j}}|^2$ and $|{h_{ie_j}}|^2$ are available, the ``best" antenna could be determined at the source by using (7).

\subsection{SAS Scheme}
As aforementioned, the OAS scheme requires that the global CSI of both the main and wiretap channels is available at the source node. However, in some cases where the eavesdropper's CSI is unavailable, the OAS scheme cannot work properly. To this end, this subsection presents a so-called suboptimal antenna selection (SAS) scheme, in which the ``best" transmit antenna would be the one that maximizes the channel capacity of the source-destination transmission, instead of maximizing the secrecy capacity. Therefore, the antenna selection criterion in SAS scheme is expressed as
\begin{equation}\label{equa8}
\begin{split}
b = \arg \mathop {\max }\limits_{i \in {\mathcal{S}}} {R_{id}}
= \arg \mathop {\max }\limits_{i \in {\mathcal{S}}} \sum\limits_{j = 1}^{{N_d}} {|{h_{id_j}}{|^2}},
\end{split}
\end{equation}
where $b$ denotes the selected ``best" transmit antenna. Meanwhile, considering the SAS scheme, the maximal achievable rate at the eavesdropper is given by
\begin{equation}\label{equa9}
R^{\textrm{SAS}}_{be}= \log_2(1+\sum\limits_{j = 1}^{{N_e}} {\frac{{P|{h_{be_j}}{|^2}}}{{\sigma _n^2}}}).
\end{equation}
We now complete the signal modeling of the STT, OAS and SAS schemes.

\section{{{Performance Analysis}}}
{{In this section, we carry out performance analysis for the STT, OAS and SAS schemes over Rayleigh fading channels. We first derive closed-form expressions of the probability of zero secrecy capacity for these three schemes. Next, the generalized secrecy diversity analysis is conducted for the sake of providing an insight into the impact of the number of antennas on the probability of zero secrecy capacity in high MER region.}}

\subsection{Probability of Zero Secrecy Capacity}
As discussed above, if the capacity of the main channel falls below that of the wiretap channel, the secrecy capacity becomes zero and thus an event of zero secrecy capacity happens. Accordingly, the probability of zero secrecy capacity is defined as follows.\\
\textbf{Definition 1}: \emph{Letting $R_m$ and $R_w$, respectively, denote the achievable rates of the main channel and wiretap channel, then the probability of zero secrecy capacity is obtained as}
\begin{equation}\label{equa10}
P_{{\textrm{zeroSC}}} = \Pr \left\{ {{R_{m}} < {R_{w}}} \right\}.
\end{equation}


\subsubsection{STT Scheme}
Let us first analyze the probability of zero secrecy capacity for the STT scheme. By using (2) and (3), the probability of zero secrecy capacity for the STT scheme is given by
\begin{equation}\label{equa11}
\begin{split}
P_{{\textrm{zeroSC}}}^{\textrm{STT}} &= \Pr \left\{ {{R^{\textrm{STT}}_{sd}} < {R^{\textrm{STT}}_{se}}} \right\} \\
&= \Pr \left\{ {\sum\limits_{i = 1}^M {\left[ {\sum\limits_{j = 1}^{{N_d}} {|{h_{id_j}}{|^2}}  - \sum\limits_{j = 1}^{{N_e}} {|{h_{ie_j}}{|^2}} } \right]}  < 0} \right\}.
\end{split}
\end{equation}
For simplicity, we consider the case, where the fading coefficients of all main links $|h_{id_j}|^2$ ($i \in {\cal{S}}, j \in {\cal{D}}$) are independent and identically distributed (i.i.d.) RVs with the same average channel gain denoted by $\sigma^2_{sd}=E\left(|h_{id_j}|^2\right)$. Moreover, the fading coefficients of all wiretap links $|h_{ie_j}|^2$ ($i \in {\cal{S}}, j \in {\cal{E}}$) are also assumed to be i.i.d. RVs with an average channel gain denoted by $\sigma^2_{se}=E\left(|h_{ie_j}|^2\right)$. For notational convenience, let $\lambda_{de}$ denote the ratio of $\sigma^2_{sd}$ to $\sigma^2_{se}$, i.e., $\lambda_{de}=\sigma^2_{sd}/\sigma^2_{se}$ that is referred to as the MER throughout this paper. Since $|h_{id_j}|^2$ and $|h_{ie_j}|^2$ are assumed to be i.i.d. with exponential distribution, then $\sum\limits_{i = 1}^M\sum\limits_{j = 1}^{{N_d}} {|{h_{id_j}}{|^2}}  $ and $\sum\limits_{i = 1}^M\sum\limits_{j = 1}^{{N_e}} {|{h_{ie_j}}{|^2}}$ are two independent Gamma random variables. Using [32], a closed-form solution to (11) is obtained as
\begin{equation}\label{equa12}
\begin{split}
P_{{\textrm{zeroSC}}}^{\textrm{STT}} =& {(1 + {\lambda _{de}})^{1 - M{N_d} - M{N_e}}}\\
&\times\sum\limits_{k = 0}^{M{N_e}-1} {\left( \begin{array}{l}
 M{N_d} + M{N_e} - 1 \\
 \quad\quad\quad k \\
 \end{array} \right)\lambda _{de}^k}.
\end{split}
\end{equation}
In addition, considering $\sum\limits_{i = 1}^M {\left[ {\sum\limits_{j = 1}^{{N_d}} {|{h_{id_j}}{|^2}}  - \sum\limits_{j = 1}^{{N_e}} {|{h_{ie_j}}{|^2}} } \right]}  \le M \cdot \mathop {\max }\limits_{i \in {\mathcal{S}}} \left[ {\sum\limits_{j = 1}^{{N_d}} {|{h_{id_j}}{|^2}}  - \sum\limits_{j = 1}^{{N_e}} {|{h_{ie_j}}{|^2}} } \right]$, we obtain a lower bound on the probability of zero secrecy capacity of (11) as given by
\begin{equation}\label{equa13}
\begin{split}
P_{{\textrm{zeroSC}}}^{\textrm{STT}} &\ge \Pr \left\{ {M\mathop {\max }\limits_{i \in {\mathcal{S}}} \left[ {\sum\limits_{j = 1}^{{N_d}} {|{h_{id_j}}{|^2}}  - \sum\limits_{j = 1}^{{N_e}} {|{h_{ie_j}}{|^2}} } \right] < 0} \right\}\\
&= \prod\limits_{i = 1}^M {\Pr \left\{ {\sum\limits_{j = 1}^{{N_d}} {|{h_{id_j}}{|^2}}  < \sum\limits_{j = 1}^{{N_e}} {|{h_{ie_j}}{|^2}} } \right\}}.
\end{split}
\end{equation}

It is worth mentioning that the lower bound as given in (13) is an exact probability of zero secrecy capacity for the OAS scheme, as will be shown in the following subsection. This theoretically shows that the probability of zero secrecy capacity for the proposed OAS scheme is strictly lower than that of the conventional STT scheme, showing the security benefit of using the OAS scheme.

\subsubsection{OAS Scheme}
In this subsection, we analyze the probability of zero secrecy capacity for the proposed OAS scheme. Denoting the maximal achievable rates of the transmission from the ``best" transmit antenna to the destination and to the eavesdropper by $R^{\textrm{OAS}}_{bd}$ and $R^{\textrm{OAS}}_{be}$, respectively, we can express the probability of zero secrecy capacity for the OAS scheme as
\begin{equation}\label{equa14}
P_{{\textrm{zeroSC}}}^{\textrm{OAS}} = \Pr \left\{ {{R^{\textrm{OAS}}_{bd}} < {R^{\textrm{OAS}}_{be}}} \right\},
\end{equation}
where ${{R^{\textrm{OAS}}_{bd}}}$ and ${{R^{\textrm{OAS}}_{be}}}$ are given by
\begin{equation}\label{equa15}
{{R^{\textrm{OAS}}_{bd}}}={\log _2}(1 + \sum\limits_{j = 1}^{{N_d}} {\frac{{|{h_{bd_j}}{|^2}P}}{{\sigma _n^2}}} ),
\end{equation}
and
\begin{equation}\label{equa16}
{{R^{\textrm{OAS}}_{be}}}={\log _2}(1 + \sum\limits_{j = 1}^{{N_e}} {\frac{{|{h_{be_j}}{|^2}P}}{{\sigma _n^2}}} ).
\end{equation}
Substituting ${{R^{\textrm{OAS}}_{bd}}}$ and ${{R^{\textrm{OAS}}_{be}}}$ from (15) and (16) into (14) yields
\begin{equation}\label{equa17}
P_{{\textrm{zeroSC}}}^{\textrm{OAS}} = \Pr \left\{\dfrac{{1 + \sum\limits_{j = 1}^{{N_d}} {\frac{{|{h_{bd_j}}{|^2}P}}{{\sigma _n^2}}} }}{{1 + \sum\limits_{j = 1}^{{N_e}} {\frac{{|{h_{be_j}}{|^2}P}}{{\sigma _n^2}}} }} < 1\right\}.
\end{equation}
Combining (7) and (17) gives
\begin{equation}\label{equa18}
P_{{\textrm{zeroSC}}}^{\textrm{OAS}} = \Pr \left\{ {\mathop {\max }\limits_{i \in {\mathcal{S}}} \left( {\frac{{1 + \sum\limits_{j = 1}^{{N_d}} {\frac{{|{h_{id_j}}{|^2}P}}{{\sigma _n^2}}} }}{{1 + \sum\limits_{j = 1}^{{N_e}} {\frac{{|{h_{ie_j}}{|^2}P}}{{\sigma _n^2}}} }}} \right) < 1} \right\},
\end{equation}
which can be further obtained as
\begin{equation}\label{equa19}
\begin{split}
P_{{\textrm{zeroSC}}}^{\textrm{OAS}} &= \prod\limits_{i = 1}^M {\Pr \left\{ {\frac{{1 + \sum\limits_{j = 1}^{{N_d}} {\frac{{|{h_{id_j}}{|^2}P}}{{\sigma _n^2}}} }}{{1 + \sum\limits_{j = 1}^{{N_e}} {\frac{{|{h_{ie_j}}{|^2}P}}{{\sigma _n^2}}} }} < 1} \right\}} \\
&= \prod\limits_{i = 1}^M {\Pr \left\{ {\sum\limits_{j = 1}^{{N_d}} {|{h_{id_j}}{|^2}}  < \sum\limits_{j = 1}^{{N_e}} {|{h_{ie_j}}{|^2}} } \right\}},
\end{split}
\end{equation}
which is exactly the same as the STT scheme's lower bound on the probability of zero secrecy capacity as given by (13), confirming the security advantage of the OAS scheme over the STT scheme. Assume that the fading coefficients $|h_{id_j}|^2$ and $|h_{ie_j}|^2$ are i.i.d. exponential RVs with respective average channel gains $\sigma^2_{sd}$ and $\sigma^2_{se}$. Following [32], we obtain a closed-form expression of (19) as
\begin{equation}\label{equa20}
\begin{split}
P_{{\textrm{zeroSC}}}^{\textrm{OAS}} =& {(1 + {\lambda _{de}})^{M(1 - {N_d} - {N_e})}}\\
&\times\left[\sum\limits_{k = 0}^{{N_e}-1} {\left( \begin{array}{l}
 {N_d} + {N_e} - 1 \\
 \quad\quad k \\
 \end{array} \right)\lambda _{de}^k}\right]^M.
\end{split}
\end{equation}

\subsubsection{SAS Scheme}
This subsection presents the probability of zero secrecy capacity analysis of the SAS scheme. As shown in (8), the SAS scheme attempts to maximize the capacity of the main channel spanning from source to destination. Hence, using (8) and (9), we obtain the probability of zero secrecy capacity for the SAS scheme as
\begin{equation}\label{equa21}
\begin{split}
P_{{\textrm{zeroSC}}}^{\textrm{SAS}}& = \Pr \left\{ {{R^{\textrm{SAS}}_{bd}} < {R^{\textrm{SAS}}_{be}}} \right\}\\
&=\Pr \left\{ { \sum\limits_{j=1}^{N_d} {|{h_{bd_j}}{|^2}}  <  \sum\limits_{j=1}^{N_e} {|{h_{be_j}}{|^2}}} \right\}\\
&=\sum\limits_{m = 1}^M {\Pr \left\{ { \sum\limits_{j=1}^{N_d} {|{h_{md_j}}{|^2}}  <  \sum\limits_{j=1}^{N_e} {|{h_{me_j}}{|^2}}},b=m \right\}},
\end{split}
\end{equation}
where $b$ denotes the ``best" antenna determined by (8) and the last equation is obtained by using the law of total probability. From (8), an event $b=m$ means
\begin{equation}\label{equa22}
\sum\limits_{j = 1}^{{N_d}} {|{h_{m{d_j}}}{|^2}}  > \mathop {\max }\limits_{i \in {\cal{S}},i \ne m} \sum\limits_{j = 1}^{{N_d}} {|{h_{i{d_j}}}{|^2}},
\end{equation}
substituting which into (21) gives
\begin{equation}\label{equa23}
\begin{split}
P_{{\textrm{zeroSC}}}^{\textrm{SAS}}=\sum\limits_{m = 1}^M {\Pr \left\{
\begin{split}
 &{ \sum\limits_{j=1}^{N_e}{|{h_{me_j}}{|^2}}}>\sum\limits_{j=1}^{N_d} {|{h_{md_j}}{|^2}},\\
 &\mathop {\max }\limits_{i \in {\cal{S}},i \ne m} \sum\limits_{j = 1}^{{N_d}} {|{h_{i{d_j}}}{|^2}}<\sum\limits_{j = 1}^{{N_d}} {|{h_{m{d_j}}}{|^2}}
\end{split}
\right \}}.
\end{split}
\end{equation}
For simplicity, we consider that the fading coefficients ${h_{i{e_j}}}(i \in {\mathcal{S}},j \in {\mathcal{E}})$ of wiretap links from $M$ transmit antennas (at source) to $N_e$ receive antennas (at eavesdropper) are i.i.d. In this way, RV $\sum\limits_{j = 1}^{{N_e}} {|{h_{m{e_j}}}{|^2}}$ follows the same distribution for different transmit antenna $m$ ($1\le m \le M$). Thus, (23) can be further simplified to
\begin{equation}\label{equa24}
P_{{\textrm{zeroSC}}}^{\textrm{SAS}} = {\Pr \left\{ {\mathop {\max }\limits_{i \in {\mathcal{S}}} \sum\limits_{j = 1}^{{N_d}} {|{h_{i{d_j}}}{|^2}}  < \sum\limits_{j = 1}^{{N_e}} {|{h_{m{e_j}}}{|^2}} } \right\}},
\end{equation}
which can be used to numerically calculate the probability of zero secrecy capacity for the SAS scheme. {As shown in (11), (19) and (23), the probabilities of zero secrecy capacity for the STT, OAS and SAS schemes are only related to the main channel $|h_{id_j}|^2$ and wiretap channel $|h_{ie_j}|^2$, which has nothing to do with the SNR. This means that increasing the transmit power cannot improve the wireless security in terms of the probability of zero secrecy capacity.}

\subsection{Generalized Secrecy Diversity}
In what follows, we are focused on the secrecy diversity analysis for characterizing an asymptotic behavior on the probability of zero secrecy capacity in high MER region. In [23] and [24], the conventional secrecy diversity is defined as
\begin{equation}\label{equa25}
d =  - \mathop {\lim }\limits_{{\textrm{SNR}}_d \to \infty }\frac {\log {P_{out}}({\textrm{SNR}}_d)}{\log {\textrm{SNR}_d}},
\end{equation}
where ${\textrm{SNR}}_d$ represents the received SNR at the destination and $P_{out}({\textrm{SNR}}_d)$ represents the secrecy outage probability. One can observe that the conventional SNR based secrecy diversity definition is not applicable here, since the probability of zero secrecy capacity is independent of the SNR and only relates to the main and wiretap channels, as shown in (11), (19) and (23). To this end, we consider the use of our generalized secrecy diversity [19]-[21] to characterize an asymptotic behavior on the probability of zero secrecy capacity in high MER region. In our generalized secrecy diversity, an asymptotic probability of zero secrecy capacity is characterized as the ratio between the average gains of the main channel and the eavesdropper's wiretap channel (i.e. MER) tends to infinity. Therefore, our generalized secrecy diversity is defined as follows.\\
\textbf{Definition 2}: \emph{Denoting $\lambda_{de}=\sigma^2_{sd}/\sigma^2_{se}$, the generalized secrecy diversity is given by the asymptotic ratio of the probability of zero secrecy capacity to MER $\lambda_{de}$, yielding}
\begin{equation} \label{equa26}
{d} = -\mathop {\lim }\limits_{{\lambda _{de}} \to \infty } \dfrac{{\log (P_{{\textrm{zeroSC}}})}}{{\log ({\lambda _{de}})}}.
\end{equation}
{It is pointed out that the MER $\lambda_{de}$ possibly approaches infinity by mitigating the eavesdropper's received signal using an advanced signal processing technique, such as beamforming. To be specific, when the beamforming is adopted, the source node could transmit its signal in a particular direction to the destination, so that the main channel experiences constructive interference, whereas destructive interference is encountered in the wiretap channel, resulting in a high MER.}

\subsubsection{STT Scheme}
In this subsection, the secrecy diversity analysis of STT scheme is presented. Using (26), the secrecy diversity of STT scheme is given by
\begin{equation}\label{equa27}
{d_{{\textrm{STT}}}} = -\mathop {\lim }\limits_{{\lambda _{de}} \to \infty } \dfrac{{\log (P_{{\textrm{zeroSC}}}^{{\textrm{STT}}})}}{{\log ({\lambda _{de}})}},
\end{equation}
where ${P_{{\textrm{zeroSC}}}^{{\textrm{STT}}}}$ is given by (11). Using the inequalities of $\sum\limits_{i = 1}^M {\sum\limits_{j = 1}^{{N_d}} {|{h_{i{d_j}}}{|^2}} }  \le M{N_d}\mathop {\max }\limits_{i \in {\mathcal{S}},j \in {\mathcal{D}}} |{h_{i{d_j}}}{|^2}$ and $\sum\limits_{i = 1}^M {\sum\limits_{j = 1}^{{N_e}} {|{h_{i{e_j}}}{|^2}} }  \ge \mathop {\max }\limits_{i \in {\mathcal{S}},j \in {\mathcal{E}}} |{h_{i{e_j}}}{|^2}$, we can obtain a lower bound on the probability of zero secrecy capacity of STT scheme as
\begin{equation}\label{equa28}
P_{{\textrm{zeroSC}}}^{{\textrm{STT}}} \ge \Pr \left\{ {M{N_d}\mathop {\max }\limits_{i \in {\mathcal{S}},j \in {\mathcal{D}}} |{h_{i{d_j}}}{|^2} < \mathop {\max }\limits_{i \in {\mathcal{S}},j \in {\mathcal{E}}} |{h_{i{e_j}}}{|^2}} \right\},
\end{equation}
where $\mathcal{D}$ and $\mathcal{E}$ represent the sets of antennas at the destination and eavesdropper, respectively. Denoting ${X_e} = \mathop {\max }\limits_{i \in {\mathcal{S}},j \in {\mathcal{E}}} |{h_{i{e_j}}}{|^2}$ and considering that $|{h_{i{e_j}}}{|^2}$ follows exponential distribution with mean $\sigma _{i{e_j}}^2$, we can obtain the cumulative distribution function (CDF) of $X_e$ as
\begin{equation}\label{equa29}
\begin{split}
{P_{{X_e}}}(x)
= 1 + \sum\limits_{k = 1}^{{2^{M{N_e}}} - 1} {{{( - 1)}^{|{{\mathcal{A}}_k}|}}\exp ( - \sum\limits_{i,j \in {{\mathcal{A}}_k}} {\frac{x}{{\sigma _{i{e_j}}^2}}} )},
\end{split}
\end{equation}
for ${\textrm{ }}x \ge 0 $; otherwise ${P_{{X_e}}}(x)=0$ for ${\textrm{ }}x < 0 $, where ${\mathcal{A}}_k$ represents the $k$-th non-empty sub-collection of $MN_e$ elements of term $\exp ( - \frac{x}{{\sigma _{i{e_j}}^2}})$ for $({i \in {\mathcal{S}},j \in {\mathcal{E}}})$, and $|{\mathcal{A}}_k|$ is the number of the elements in ${\mathcal{A}}_k$. From (29), the probability density function (PDF) of $X_e$ can be derived as
\begin{equation}\label{equa30}
{p_{{X_e}}}(x) = \sum\limits_{k = 1}^{{2^{M{N_e}}} - 1} {\sum\limits_{i,j \in {{\mathcal{A}}_k}} {\frac{{{{( - 1)}^{|{{\mathcal{A}}_k}| + 1}}}}{{\sigma _{i{e_j}}^2}}} \exp ( - \sum\limits_{i,j \in {{\mathcal{A}}_k}} {\frac{x}{{\sigma _{i{e_j}}^2}}} )},
\end{equation}
for $x \ge 0$. Using (28) and (30), we have
\begin{equation}\label{equa31}
P_{{\textrm{zeroSC}}}^{{\textrm{STT}}} \ge \int_{-\infty}^\infty  {\prod\limits_{i \in {\mathcal{S}},j \in {\mathcal{D}}} {[1 - \exp ( - \frac{x}{{M{N_d}\sigma _{i{d_j}}^2}})]}  {p_{{X_e}}}(x)dx},
\end{equation}
where ${{p_{{X_e}}}(x)}$ is given in (30).\\
\textbf{Theorem 1:} \emph{Considering a random variable $x$ with PDF given by (30), then the following equation holds for ${\lambda _{de}} \to \infty $}
\begin{equation}\nonumber
1 - \exp ( - \dfrac{x}{{M{N_d}\sigma _{i{d_j}}^2}}) \buildrel 1 \over =  \dfrac{x}{{M{N_d}\sigma _{i{d_j}}^2}},
\end{equation}
\emph{where $\buildrel 1 \over = $ represents an equality with probability 1.}\\
\textbf{Proof:} \emph{See Appendix A}.

Substituting (30) into (31) and using Theorem 1, we obtain
\begin{equation}\label{equa32}
\begin{split}
P_{{\textrm{zeroSC}}}^{{\textrm{STT}}} &\ge \sum\limits_{k = 1}^{{2^{M{N_e}}} - 1} {{(\frac{1}{{M{N_d}}})}^{M{N_d}}}\prod\limits_{i \in {\mathcal{S}},j \in {\mathcal{D}}} {\frac{1}{{\sigma _{i{d_j}}^2}}} \\
& \times\int_0^\infty  {\sum\limits_{i,j \in {{\mathcal{A}}_k}} {\frac{{{{( - 1)}^{|{{\mathcal{A}}_k}| + 1}}}{x^{M{N_d}}}}{{\sigma _{i{e_j}}^2}}} \exp ( - \sum\limits_{i,j \in {{\mathcal{A}}_k}} {\frac{x}{{\sigma _{i{e_j}}^2}}} )dx}  \\
& = \sum\limits_{k = 1}^{{2^{M{N_e}}} - 1} {{{( - 1)}^{|{{\mathcal{A}}_k}| + 1}}\frac{{(M{N_d})!\prod\limits_{i \in {\mathcal{S}},j \in {\mathcal{D}}} {\frac{1}{{\sigma _{i{d_j}}^2}}} }}{{{{(\sum\limits_{i,j \in {{\mathcal{A}}_k}} {\frac{{M{N_d}}}{{\sigma _{i{e_j}}^2}}} )}^{M{N_d}}}}}},
\end{split}
\end{equation}
for ${\lambda _{de}} \to \infty $. By denoting $\sigma _{i{d_j}}^2 = {\alpha _{i{d_j}}}\sigma _{sd}^2$ and $\sigma _{i{e_j}}^2 = {\alpha _{i{e_j}}}\sigma _{se}^2$, (32) can be rewritten as
\begin{equation}\label{equa33}
P_{{\textrm{zeroSC}}}^{{\textrm{STT}}} \ge \sum\limits_{k = 1}^{{2^{M{N_e}}} - 1} {\frac{{{( - 1)}^{|{{\mathcal{A}}_k}| + 1}}{(M{N_d})! }}{{{{(\sum\limits_{i,j \in {{\mathcal{A}}_k}} {\frac{{M{N_d}}}{{{\alpha _{i{e_j}}}}}} )}^{M{N_d}}}}\prod\limits_{i \in {\mathcal{S}},j \in {\mathcal{D}}} {{{{\alpha _{i{d_j}}}}}}}}  {\left( {\frac{1}{{{\lambda _{de}}}}} \right)^{M{N_d}}}
\end{equation}
where ${\lambda _{de}} = \sigma _{sd}^2/\sigma _{se}^2$. Thus, substituting (33) into (27) gives
\begin{equation}\label{equa34}
d_{{\textrm{STT}}}\le M{N_d}.
\end{equation}

In addition, considering $\sum\limits_{i = 1}^M {\sum\limits_{j = 1}^{{N_d}} {|{h_{i{d_j}}}{|^2}} }  \ge \mathop {\max }\limits_{i \in {\mathcal{S}},j \in {\mathcal{D}}} |{h_{i{d_j}}}{|^2}$ and $\sum\limits_{i = 1}^M {\sum\limits_{j = 1}^{{N_e}} {|{h_{i{e_j}}}{|^2}} }  \le M{N_e}\mathop {\max }\limits_{i \in {\mathcal{S}},j \in {\mathcal{E}}} |{h_{i{e_j}}}{|^2}$, we obtain an upper bound on the probability of zero secrecy capacity of STT scheme as
\begin{equation}\label{equa35}
P_{{\textrm{zeroSC}}}^{{\textrm{STT}}} \le \Pr \left\{ {\mathop {\max }\limits_{i \in {\mathcal{S}},j \in {\mathcal{D}}} |{h_{i{d_j}}}{|^2} < M{N_e}\mathop {\max }\limits_{i \in {\mathcal{S}},j \in {\mathcal{E}}} |{h_{i{e_j}}}{|^2}} \right\}.
\end{equation}
Using (30) and letting ${\lambda _{de}} \to \infty $, (35) can be further obtained expressed in closed-form as
\begin{equation}\label{equa36}
\begin{split}
&P_{{\textrm{zeroSC}}}^{{\textrm{STT}}} \le \int_{-\infty}^\infty  {\prod\limits_{i \in {\mathcal{S}},j \in {\mathcal{D}}} {[1 - \exp ( - \frac{{M{N_e}x}}{{\sigma _{i{d_j}}^2}})]}  {p_{{X_e}}}(x)dx} \\
&= \sum\limits_{k = 1}^{{2^{M{N_e}}} - 1} {{{( - 1)}^{|{{\mathcal{A}}_k}| + 1}}\frac{{{{(M{N_e})}^{M{N_d}}}(M{N_d})! }}{{{{(\sum\limits_{i,j \in {{\mathcal{A}}_k}} {\frac{1}{{{\alpha _{i{e_j}}}}}} )}^{M{N_d}}}}\prod\limits_{i \in {\mathcal{S}},j \in {\mathcal{D}}} {{{{\alpha _{i{d_j}}}}}} }}\\
&\quad\times {\left( {\frac{1}{{{\lambda _{de}}}}} \right)^{M{N_d}}},
\end{split}
\end{equation}
where the second equation is obtained by using $1 - \exp ( - \frac{{M{N_e}x}}{{\sigma _{i{d_j}}^2}}) \buildrel 1 \over = \frac{{M{N_e}x}}{{\sigma _{i{d_j}}^2}}$ with ${\lambda _{de}} \to \infty $. Substituting (36) into (27) gives
\begin{equation}\label{equa37}
d_{{\textrm{STT}}}\ge M{N_d}.
\end{equation}
Therefore, by combining (34) and (37), the secrecy diversity of STT scheme is readily obtained as
\begin{equation}\label{equa38}
d_{{\textrm{STT}}}= M{N_d}.
\end{equation}

One can observe from (38) that the secrecy diversity of STT scheme is the product of the number of transmit antennas $M$ and that of receive antennas at the destination $N_d$, which is independent of the number of eavesdropper's antennas $N_e$. This implies that the secrecy diversity of STT scheme is insusceptible to the eavesdropper. More specifically, although increasing the number of eavesdropper's antennas would definitely degrade the probability of zero secrecy capacity, it will not affect the speed at which the probability of zero secrecy capacity decreases as $\lambda_{de} \to \infty$. Moreover, as $M $ and $N_d$ increase, the secrecy diversity $MN_d$ of STT scheme increases accordingly, meaning that increasing the number of antennas at the source and destination can significantly improve the speed at which the probability of zero secrecy capacity decreases as $\lambda_{de} \to \infty$.

\subsubsection{OAS Scheme}
In this subsection, we present the secrecy diversity analysis of proposed OAS scheme. Similarly to (27), the secrecy diversity of OAS scheme is given by
\begin{equation}\label{equa39}
{d_{{\textrm{OAS}}}} = -\mathop {\lim }\limits_{{\lambda _{de}} \to \infty } \dfrac{{\log (P_{{\textrm{zeroSC}}}^{{\textrm{OAS}}})}}{{\log ({\lambda _{de}})}},
\end{equation}
where ${P_{{\textrm{zeroSC}}}^{{\textrm{OAS}}}}$ is given by (19). Considering inequalities $\sum\limits_{j = 1}^{{N_d}} {|{h_{i{d_j}}}{|^2}}  \le {N_d}\mathop {\max }\limits_{j \in {\mathcal{D}}} |{h_{i{d_j}}}{|^2}$ and $\sum\limits_{j = 1}^{{N_e}} {|{h_{i{e_j}}}{|^2}}  \ge \mathop {\max }\limits_{j \in {\mathcal{E}}} |{h_{i{e_j}}}{|^2}$, a lower bound on the probability of zero secrecy capacity of the OAS scheme is obtained as
\begin{equation}\label{equa40}
P_{{\textrm{zeroSC}}}^{{\textrm{OAS}}} \ge \prod\limits_{i = 1}^M {\Pr \left\{ {{N_d}\mathop {\max }\limits_{j \in {\mathcal{D}}} |{h_{i{d_j}}}{|^2} < \mathop {\max }\limits_{j \in {\mathcal{E}}} |{h_{i{e_j}}}{|^2}} \right\}}.
\end{equation}
Denoting ${Y_e} = \mathop {\max }\limits_{j \in {\mathcal{E}}} |{h_{i{e_j}}}{|^2}$, we can easily derive the PDF of $Y_e$ as
\begin{equation}\label{equa41}
{p_{{Y_e}}}(y) =
\sum\limits_{k = 1}^{{2^{{N_e}}} - 1} {\sum\limits_{j \in {{\mathcal{B}}_k}} {\dfrac{{{{( - 1)}^{|{{\mathcal{B}}_k}| + 1}}}}{{\sigma _{i{e_j}}^2}}} \exp ( - \sum\limits_{j \in {{\mathcal{B}}_k}} {\dfrac{y}{{\sigma _{i{e_j}}^2}}} )} ,
\end{equation}
for ${\textrm{ }}y \ge 0$; otherwise ${p_{{Y_e}}}(y) = 0$ for ${\textrm{ }}y < 0$, where ${\mathcal{B}}_k$ is the $k$-th non-empty subset of $N_e$ receive antennas at eavesdropper and $|{\mathcal{B}}_k|$ is the number of elements of set ${\mathcal{B}}_k$. Using (40) and (41), we have
\begin{equation}\label{equa42}
\begin{split}
P_{{\textrm{zeroSC}}}^{{\textrm{OAS}}} \ge &\prod\limits_{i = 1}^M \int_0^\infty \prod\limits_{j = 1}^{{N_d}} {[1 - \exp ( - \frac{y}{{{N_d}\sigma _{i{d_j}}^2}})]}  \\
&\quad\times\sum\limits_{k = 1}^{{2^{{N_e}}} - 1} {\sum\limits_{j \in {{\mathcal{B}}_k}} {\frac{{{{( - 1)}^{|{{\mathcal{B}}_k}| + 1}}}}{{\sigma _{i{e_j}}^2}}} \exp ( - \sum\limits_{j \in {{\mathcal{B}}_k}} {\frac{y}{{\sigma _{i{e_j}}^2}}} )} dy  .
\end{split}
\end{equation}
Considering ${\lambda _{de}} \to \infty $, we have ${1 - \exp ( - \frac{y}{{N_d\sigma _{i{d_j}}^2}})}\buildrel 1 \over ={\frac{y}{{N_d\sigma _{i{d_j}}^2}}}$ by using Theorem 1. Substituting this result into (42) yields
\begin{equation}\label{equa43}
\begin{split}
P_{{\textrm{zeroSC}}}^{{\textrm{OAS}}} &\ge \prod\limits_{i = 1}^M \sum\limits_{k = 1}^{{2^{{N_e}}} - 1} {{(\frac{1}{{{N_d}}})}^{{N_d}}}\prod\limits_{j = 1}^{{N_d}} {\frac{1}{{\sigma _{i{d_j}}^2}}}\\
&\quad\times\int_0^\infty  {\sum\limits_{j \in {{\mathcal{B}}_k}} {\frac{{{{( - 1)}^{|{{\mathcal{B}}_k}| + 1}}{y^{{N_d}}}}}{{\sigma _{i{e_j}}^2}}} \exp ( - \sum\limits_{j \in {{\mathcal{B}}_k}} {\frac{y}{{\sigma _{i{e_j}}^2}}} )dy}   \\
&= \prod\limits_{i = 1}^M {\left[ {\sum\limits_{k = 1}^{{2^{{N_e}}} - 1} {{{( - 1)}^{|{{\mathcal{B}}_k}| + 1}}\frac{{{N_d}!\prod\limits_{j = 1}^{{N_d}} {\frac{1}{{{\alpha _{i{d_j}}}}}} }}{{{{(\sum\limits_{j \in {{\mathcal{B}}_k}} {\frac{{{N_d}}}{{{\alpha _{i{e_j}}}}}} )}^{{N_d}}}}}} } \right]}  {(\frac{1}{{{\lambda _{de}}}})^{M{N_d}}},
\end{split}
\end{equation}
for ${\lambda _{de}} \to \infty $. Thus, substituting (43) into (39) gives
\begin{equation}\label{equa44}
d_{{\textrm{OAS}}} \le M{N_d}.
\end{equation}

In addition, by using inequalities $\sum\limits_{j = 1}^{{N_d}} {|{h_{i{d_j}}}{|^2}}  \ge \mathop {\max }\limits_{j \in {\mathcal{D}}} |{h_{i{d_j}}}{|^2}$ and $\sum\limits_{j = 1}^{{N_e}} {|{h_{i{e_j}}}{|^2}}  \le {N_e}\mathop {\max }\limits_{j \in {\mathcal{E}}} |{h_{i{e_j}}}{|^2}$, an upper bound on the probability of zero secrecy capacity of OAS scheme is given by
\begin{equation}\label{equa45}
P_{{\textrm{zeroSC}}}^{{\textrm{OAS}}} \le \prod\limits_{i = 1}^M {\Pr \left\{ {\mathop {\max }\limits_{j \in {\mathcal{D}}} |{h_{i{d_j}}}{|^2} < {N_e}\mathop {\max }\limits_{j \in {\mathcal{E}}} |{h_{i{e_j}}}{|^2}} \right\}},
\end{equation}
which can be further expressed as (46) for ${\lambda _{de}} \to \infty $ at the top of the following page.
\begin{figure*}
\begin{equation}\label{equa46}
\begin{split}
P_{{\textrm{zeroSC}}}^{{\textrm{OAS}}}&\le \prod\limits_{i = 1}^M {\Pr \left\{ {\mathop {\max }\limits_{j \in {\mathcal{D}}} |{h_{i{d_j}}}{|^2} < {N_e}\mathop {\max }\limits_{j \in {\mathcal{E}}} |{h_{i{e_j}}}{|^2}} \right\}} \\
&= \prod\limits_{i = 1}^M {\int_0^\infty  {\prod\limits_{j = 1}^{{N_d}} {[1 - \exp ( - \frac{{{N_e}y}}{{\sigma _{i{d_j}}^2}})]}  \sum\limits_{k = 1}^{{2^{{N_e}}} - 1} {\sum\limits_{j \in {{\mathcal{B}}_k}} {\frac{{{{( - 1)}^{|{{\mathcal{B}}_k}| + 1}}}}{{\sigma _{i{e_j}}^2}}} \exp ( - \sum\limits_{j \in {{\mathcal{B}}_k}} {\frac{y}{{\sigma _{i{e_j}}^2}}} )} dy} } \\
&= \prod\limits_{i = 1}^M {\sum\limits_{k = 1}^{{2^{{N_e}}} - 1} {{{({N_e})}^{{N_d}}}\prod\limits_{j = 1}^{{N_d}} {\frac{1}{{\sigma _{i{d_j}}^2}}} \int_0^\infty  {\sum\limits_{j \in {{\mathcal{B}}_k}} {\frac{{{{( - 1)}^{|{{\mathcal{B}}_k}| + 1}}{y^{{N_d}}}}}{{\sigma _{i{e_j}}^2}}} \exp ( - \sum\limits_{j \in {{\mathcal{B}}_k}} {\frac{y}{{\sigma _{i{e_j}}^2}}} )dy} } } \\
&= \prod\limits_{i = 1}^M {\left[ {\sum\limits_{k = 1}^{{2^{{N_e}}} - 1} {{{( - 1)}^{|{{\mathcal{B}}_k}| + 1}}{{({N_e})}^{{N_d}}}\frac{{{N_d}!\prod\limits_{j = 1}^{{N_d}} {\frac{1}{{{\alpha _{i{d_j}}}}}} }}{{{{(\sum\limits_{j \in {{\mathcal{B}}_k}} {\frac{1}{{{\alpha _{i{e_j}}}}}} )}^{{N_d}}}}}} } \right]}  {(\frac{1}{{{\lambda _{de}}}})^{M{N_d}}}
\end{split}
\end{equation}
\end{figure*}
Substituting (46) into (39) gives
\begin{equation}\label{equa47}
d_{{\textrm{OAS}}} \ge M{N_d}.
\end{equation}
Using (44) and (47), we can easily obtain the secrecy diversity of OAS scheme with the squeeze theorem as
\begin{equation}\label{equa48}
d_{{\textrm{OAS}}} = M{N_d},
\end{equation}
which shows that the OAS scheme achieves the same secrecy diversity as the STT scheme. It is pointed out that the same secrecy diversity order achieved by both the OAS and STT schemes only means that the probabilities of zero secrecy capacity of the two schemes are reduced at the same speed as $\lambda_{de} \to \infty$.

\subsubsection{SAS Scheme}
This subsection examines the secrecy diversity of SAS scheme. Similarly, the secrecy diversity of SAS scheme can be defined as
\begin{equation}\label{equa49}
{d_{{\textrm{SAS}}}} = -\mathop {\lim }\limits_{{\lambda _{de}} \to \infty } \dfrac{{\log (P_{{\textrm{zeroSC}}}^{{\textrm{SAS}}})}}{{\log ({\lambda _{de}})}},
\end{equation}
where ${P_{{\textrm{zeroSC}}}^{{\textrm{SAS}}}}$ is given by (24). Using inequalities $\mathop {\max }\limits_{i \in {\mathcal{S}}} \sum\limits_{j = 1}^{{N_d}} {|{h_{i{d_j}}}{|^2}}  \le {N_d}\mathop {\max }\limits_{i \in {\mathcal{S}},j \in {\mathcal{D}}} |{h_{i{d_j}}}{|^2}$ and $\sum\limits_{j = 1}^{{N_e}} {|{h_{m{e_j}}}{|^2}}  \ge \mathop {\max }\limits_{j \in {\mathcal{E}}} |{h_{m{e_j}}}{|^2}$, we obtain a lower bound on the probability of zero secrecy capacity of SAS scheme as
\begin{equation}\label{equa50}
P_{{\textrm{zeroSC}}}^{{\textrm{SAS}}} \ge {\Pr \left\{ {{N_d}\mathop {\max }\limits_{i \in {\mathcal{S}},j \in {\mathcal{D}}} |{h_{i{d_j}}}{|^2} < \mathop {\max }\limits_{j \in {\mathcal{E}}} |{h_{m{e_j}}}{|^2}} \right\}}.
\end{equation}
Letting ${\lambda _{de}} \to \infty $ and using (41), the preceding equation can be further calculated as
\begin{equation}\label{equa51}
\begin{split}
&P_{{\textrm{zeroSC}}}^{{\textrm{SAS}}}\ge {\Pr \left\{ {{N_d}\mathop {\max }\limits_{i \in {\mathcal{S}},j \in {\mathcal{D}}} |{h_{i{d_j}}}{|^2} < \mathop {\max }\limits_{j \in {\mathcal{E}}} |{h_{m{e_j}}}{|^2}} \right\}} \\
&=  \int_0^\infty  \prod\limits_{i \in {\mathcal{S}},j \in {\mathcal{D}}} {[1 - \exp ( - \frac{y}{{{N_d}\sigma _{i{d_j}}^2}})]} \\
&\quad \times\sum\limits_{k = 1}^{{2^{{N_e}}} - 1} {\sum\limits_{j \in {{\mathcal{B}}_k}} {\frac{{{{( - 1)}^{|{{\mathcal{B}}_k}| + 1}}}}{{\sigma _{m{e_j}}^2}}} \exp ( - \sum\limits_{j \in {{\mathcal{B}}_k}} {\frac{y}{{\sigma _{m{e_j}}^2}}} )} dy  \\
&=  {\sum\limits_{k = 1}^{{2^{{N_e}}} - 1} {{{( - 1)}^{|{{\mathcal{B}}_k}| + 1}}\frac{{(M{N_d})!\prod\limits_{i \in {\mathcal{S}},j \in {\mathcal{D}}} {(\frac{1}{{{\alpha _{i{d_j}}}}})} }}{{(\sum\limits_{j \in {{\mathcal{B}}_k}} {\frac{{{N_d}}}{{{\alpha _{m{e_j}}}}}{)^{M{N_d}}}} }}} }  {(\frac{1}{{{\lambda _{de}}}})^{M{N_d}}},
\end{split}
\end{equation}
where the third equation is obtained by using $1 - \exp ( - \frac{y}{{{N_d}\sigma _{i{d_j}}^2}}) \buildrel 1 \over = \frac{y}{{{N_d}\sigma _{i{d_j}}^2}}$ with ${\lambda _{de}} \to \infty $. Substituting (51) into (49) yields
\begin{equation}\label{equa52}
d_{{\textrm{SAS}}} \le M{N_d}.
\end{equation}
In addition, considering $\mathop {\max }\limits_{i \in {\mathcal{S}}} \sum\limits_{j = 1}^{{N_d}} {|{h_{i{d_j}}}{|^2}}  \ge \mathop {\max }\limits_{i \in {\mathcal{S}},j \in {\mathcal{D}}} |{h_{i{d_j}}}{|^2}$ and $\sum\limits_{j = 1}^{{N_e}} {|{h_{m{e_j}}}{|^2}}  \le {N_e}\mathop {\max }\limits_{j \in {\mathcal{E}}} |{h_{m{e_j}}}{|^2}$, an upper bound on the probability of zero secrecy capacity of SAS scheme is given by
\begin{equation}\label{equa53}
\begin{split}
&P_{{\textrm{zeroSC}}}^{{\textrm{SAS}}} \le  {\Pr \left\{ {\mathop {\max }\limits_{i \in {\mathcal{S}},j \in {\mathcal{D}}} |{h_{i{d_j}}}{|^2} < {N_e}\mathop {\max }\limits_{j \in {\mathcal{E}}} |{h_{m{e_j}}}{|^2}} \right\}} \\
&=  \int_0^\infty  \prod\limits_{i \in {\mathcal{S}},j \in {\mathcal{D}}} {[1 - \exp ( - \frac{{{N_e}y}}{{\sigma _{i{d_j}}^2}})]} \\
&\quad\quad\times \sum\limits_{k = 1}^{{2^{{N_e}}} - 1} {\sum\limits_{j \in {{\mathcal{B}}_k}} {\frac{{{{( - 1)}^{|{{\mathcal{B}}_k}| + 1}}}}{{\sigma _{m{e_j}}^2}}} \exp ( - \sum\limits_{j \in {{\mathcal{B}}_k}} {\frac{y}{{\sigma _{m{e_j}}^2}}} )} dy  \\
&=  {\sum\limits_{k = 1}^{{2^{{N_e}}} - 1} {\frac{{{( - 1)}^{|{{\mathcal{B}}_k}| + 1}}{{({N_e})}^{M{N_d}}}{(M{N_d})! }}{{(\sum\limits_{j \in {{\mathcal{B}}_k}} {\frac{1}{{{\alpha _{m{e_j}}}}}{)^{M{N_d}}}} }\prod\limits_{i \in {\mathcal{S}},j \in {\mathcal{D}}} {({{{\alpha _{i{d_j}}}}})}}} } {(\frac{1}{{{\lambda _{de}}}})^{M{N_d}}},
\end{split}
\end{equation}
for ${\lambda _{de}} \to \infty $. Using (49) and (53), we have
\begin{equation}\label{equa54}
d_{{\textrm{SAS}}} \ge M{N_d}.
\end{equation}
Therefore, we obtain the secrecy diversity of the SAS scheme from (52) and (54) as
\begin{equation}\label{equa55}
d_{{\textrm{SAS}}} = M{N_d},
\end{equation}
which shows that the proposed SAS scheme achieves the same secrecy diversity as the STT and OAS schemes. This means that in high MER region, the probabilities of zero secrecy capacity of STT, OAS and SAS schemes all behave as $(1/\lambda_{de})^{MN_d}$ as $\lambda_{de} \to \infty$.

\section{Numerical Results and Discussions}
\begin{figure}
  \centering
  {\includegraphics[scale=0.55]{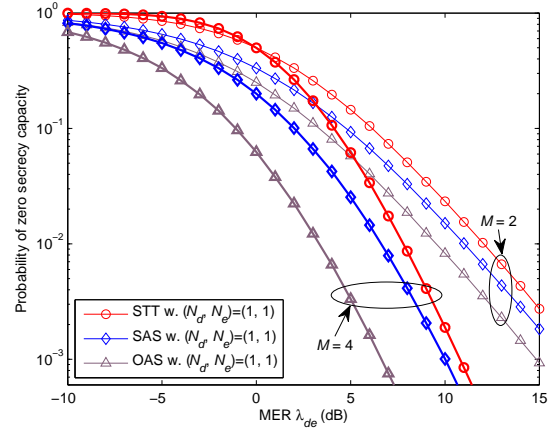}\\
  \caption{{Probability of zero secrecy capacity versus MER of the STT, SAS and OAS schemes with $\alpha_{id_j}=\alpha_{ie_j}=1$.}}\label{Fig2}}
\end{figure}

In this section, we present numerical performance results of the STT, SAS and OAS schemes in terms of the probability of zero secrecy capacity. Fig. 2 shows the probability of zero secrecy capacity versus MER of the STT, SAS and OAS schemes with $N_d=N_e=1$ and $\alpha_{id_j}=\alpha_{ie_j}=1$. From Fig. 2, one can see that as the number of transmit antennas $M$ increases from $M=2$ to $4$, the probabilities of zero secrecy capacity of the STT, SAS and OAS schemes decrease significantly, showing the security benefits of exploiting multiple transmit antennas at the source node. Fig. 2 also illustrates that for both the cases of $M=2$ and $M=4$, the OAS scheme achieves the best security performance and moreover, the SAS scheme performs better than the STT scheme in terms of the probability of zero secrecy capacity.
\begin{figure}
  \centering
  {\includegraphics[scale=0.55]{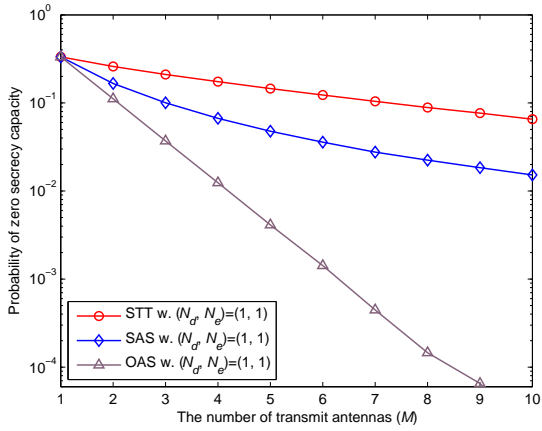}\\
  \caption{{Probability of zero secrecy capacity versus the number of transmit antennas $M$ of the STT, SAS and OAS schemes with $N_d=N_e=1$, $\lambda_{de}=3{\textrm{dB}}$, and $\alpha_{id_j}=\alpha_{ie_j}=1$.}}\label{Fig3}}
\end{figure}

Fig. 3 depicts the probability of zero secrecy capacity versus of the number of transmit antennas $M$ of the STT, SAS and OAS schemes with $N_d=N_e=1$, $\lambda_{de}=3{\textrm{dB}}$, and $\alpha_{id_j}=\alpha_{ie_j}=1$. It is shown from Fig. 3 that the OAS scheme outperforms both the SAS and STT schemes in terms of its probability of zero secrecy capacity. Moreover, as the number of transmit antennas $M$ increases, the security advantage of the OAS scheme over the SAS and STT approaches become much more significant. One can also observe from Fig. 3 that the conventional STT scheme performs worse than both the SAS and OAS schemes, showing the security benefits of using the transmit antenna selection.

\begin{figure}
  \centering
  {\includegraphics[scale=0.55]{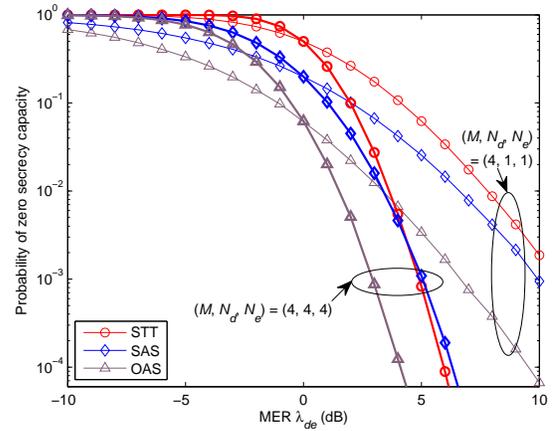}\\
  \caption{{Probability of zero secrecy capacity versus MER of the STT, SAS and OAS schemes with $M=4$, $N_d=N_e $, and $\alpha_{id_j}=\alpha_{ie_j}=1$.}}\label{Fig4}}
\end{figure}

In Figs. 2 and 3, only the source node is assumed with multiple antennas to show the performance improvement through using multiple transmit antennas. Next, we consider that the destination and eavesdroppers are also equipped with multiple antennas in evaluating the probability of zero secrecy capacity. Fig. 4 shows the probability of zero secrecy capacity versus MER of the STT, SAS and OAS schemes with $M=4$ and $\alpha_{id_j}=\alpha_{ie_j}=1$. It is seen from Fig. 4 that for both the cases of $(M, N_d, N_e)=(4,1,1)$ and $(M, N_d, N_e)=(4,4,4)$, the OAS scheme strictly achieves a better secrecy performance than both the SAS and STT schemes. However, for $(M, N_d, N_e)=(4,4,4)$, as the MER increases from $-10{\textrm{dB}}$ to $10{\textrm{dB}}$, the SAS scheme initially outperforms the STT scheme and eventually performs worse than the STT scheme in terms of probability of zero secrecy capacity.

Fig. 5 illustrates the asymptotic and exact results on the probability of zero secrecy capacity for the proposed OAS scheme by plotting (20), (43) and (46) as a function of MER $\lambda_{de}$. Specifically, the exact probability of zero secrecy capacity for the OAS scheme is obtained using (20), while the corresponding lower and upper bounds are plotted from (43) and (46), respectively. As shown in Fig. 5, upon increasing the MER, the exact probability of zero secrecy capacity falls between the lower and upper bounds. Notice that the lower and upper bounds are based on Theorem 1, which is valid for high MER only. Additionally, one can observe from Fig. 5 that in high MER region, the slopes of these three performance curves are the same, which confirms the correctness of the secrecy diversity analysis based on the squeeze theorem.

\begin{figure}
  \centering
  {\includegraphics[scale=0.55]{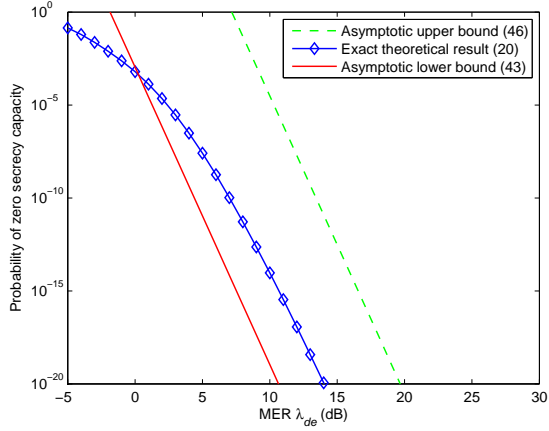}\\
  \caption{{Asymptotic and exact results on the probability of zero secrecy capacity for the proposed OAS scheme with $M=N_d=4$, $N_e=2$, and $\alpha_{id_j}=\alpha_{ie_j}=1$.}}\label{Fig5}}
\end{figure}

\section{Conclusion}
{In this paper, we explored the physical-layer security of a MIMO system comprised of one source and one destination in the presence of an eavesdropper, where each node is equipped with multiple antennas. We proposed two transmit antenna selection schemes, namely the OAS and SAS, which operate depending on whether or not the global CSI knowledge of both the main and wiretap channels is available at the source. For the purpose of performance comparison, we also considered the conventional STT scheme as a benchmark. We derived closed-form expressions of the probability of zero secrecy capacity for the STT, OAS and SAS schemes in Rayleigh fading environments. We further examined the generalized secrecy diversity of STT, SAS, and OAS schemes through an asymptotic analysis of the probability of zero secrecy capacity in high MER region. It was shown that the generalized secrecy diversity orders of STT, SAS and OAS schemes are the product of the number of antennas at the source and destination. Additionally, numerical results demonstrated that the OAS scheme outperforms both the SAS and STT schemes in terms of its probability of zero secrecy capacity, confirming the security benefits of using the optimal antenna selection against eavesdropping.}

\appendices
\section{Proof of Theorem 1}
Letting $z = \frac{x}{{M{N_d}\sigma _{i{d_j}}^2}}$ and using (30), we obtain the expected value of random variable $z$ as
\begin{equation}\nonumber
\begin{split}
E(z)
&= \int_0^\infty  \sum\limits_{k = 1}^{{2^{M{N_e}}} - 1} \sum\limits_{i,j \in {A_k}} \frac{{{{( - 1)}^{|{A_k}| + 1}}}x}{{M{N_d}\sigma _{i{d_j}}^2}{\sigma _{i{e_j}}^2}}\\
&\quad\quad\times\exp ( - \sum\limits_{i,j \in {A_k}} {\frac{x}{{\sigma _{i{e_j}}^2}}} ) dx \\
&= \sum\limits_{k = 1}^{{2^{M{N_e}}} - 1} {\frac{{{{( - 1)}^{|{A_k}| + 1}}{{(\sum\limits_{i,j \in {A_k}} {\sigma _{i{e_j}}^{ - 2}} )}^{ - 1}}}}{{M{N_d}\sigma _{i{d_j}}^2}}}.
\end{split}\label{A.1}\tag{A.1}
\end{equation}
Denoting $\sigma _{i{d_j}}^2 = {\alpha _{i{d_j}}}\sigma _{sd}^2$ and $\sigma _{i{e_j}}^2 = {\alpha _{i{e_j}}}\sigma _{se}^2$ and letting ${\lambda _{de}} \to \infty $, we have
\begin{equation}
E(z) = \sum\limits_{k = 1}^{{2^{M{N_e}}} - 1} {\frac{{{{( - 1)}^{|{A_k}| + 1}}{{(\sum\limits_{i,j \in {A_k}} {\alpha _{i{e_j}}^{ - 2}} )}^{ - 1}}}}{{M{N_d}{\alpha _{i{d_j}}}}}}  \frac{1}{{{\lambda _{de}}}},
\label{A.2}\tag{A.2}
\end{equation}
which shows that $E(z)$ tends to zero as ${\lambda _{de}} \to \infty $. Meanwhile, we can obtain the expected value of $z^2$ as
\begin{equation}
\begin{split}
E({z^2})
 = \sum\limits_{k = 1}^{{2^{M{N_e}}} - 1} {\frac{{2{{( - 1)}^{|{A_k}| + 1}}{{(\sum\limits_{i,j \in {A_k}} {\alpha _{i{e_j}}^{ - 2}} )}^{ - 2}}}}{{{{(M{N_d}\alpha _{i{d_j}}^2)}^2}}}}  {(\frac{1}{{{\lambda _{de}}}})^2}.
\end{split}\label{A.3}\tag{A.3}
\end{equation}
From (A.3), one can see that $E({z^2})$ tends to zero as ${\lambda _{de}} \to \infty $. Thus, the variance of $z$ given by $ \emph{Var}(z)=E({z^2}) - {[E(z)]^2}$ also converges to zero for ${\lambda _{de}} \to \infty $, since both $E(z)$ and $E({z^2})$ approach zero as shown in (A.2) and (A.3). Considering the fact that both mean and variance of $z$ converge to zero as ${\lambda _{de}} \to \infty $, one can conclude that random variable $z$ approaches zero with probability 1 for ${\lambda _{de}} \to \infty $. Here, we use a symbol $\buildrel 1 \over =$ to denote an equality with probability 1 as ${\lambda _{de}} \to \infty $, i.e., we write $ z \buildrel 1 \over = 0$ to represent
\begin{equation}
\Pr (\mathop {\lim }\limits_{{\lambda _{de}} \to \infty } z = 0) = 1.\label{A.4}\tag{A.4}
\end{equation}
In addition, using the Maclaurin series expansion and Cauchy's mean value theorem, we can easily obtain
\begin{equation}
1 - \exp ( - z) = z + \frac{{{z^2}}}{2}\exp ( - \theta ),\label{A.5}\tag{A.5}
\end{equation}
where $0<\theta<z$. From (A.5), we have
\begin{equation}
\mathop {\lim }\limits_{{\lambda _{de}} \to \infty } 1 - \exp ( - z) - z = \mathop {\lim }\limits_{{\lambda _{de}} \to \infty } \frac{{{z^2}}}{2}\exp ( - \theta ). \label{A.6}\tag{A.6}
\end{equation}
Similar to (A.2) and (A.3), we can prove that both mean and variance of $z^2$ converge to zero as ${{\lambda _{de}} \to \infty }$, i.e., ${{z^2}}\buildrel 1 \over = 0$. Meanwhile, considering $0< \exp ( - \theta )<1$ due to $0<\theta<z$, we obtain $\mathop {\lim }\limits_{{\lambda _{de}} \to \infty } \frac{{{z^2}}}{2}\exp ( - \theta )=0$. Substituting this result into (A.6) yields
\begin{equation}
1 - \exp ( - z) - z \buildrel 1 \over= 0.\label{A.7}\tag{A.7}
\end{equation}
Using $z = \frac{x}{{M{N_d}\sigma _{i{d_j}}^2}}$ and (A.7), we obtain
\begin{equation}
1 - \exp ( - \dfrac{x}{{M{N_d}\sigma _{i{d_j}}^2}}) \buildrel 1 \over =  \dfrac{x}{{M{N_d}\sigma _{i{d_j}}^2}},\label{A.8}\tag{A.8}
\end{equation}
for $\lambda_{de} \to \infty$. This completes the proof of Theorem 1.

\end{document}